\documentclass[10pt,preprint]{aastex}
\usepackage{natbib}

\def\d{{\:\rm d}} 
\newcommand{\Tbb}{\ensuremath{\mathit{T_{bb}}}}
\newcommand{\Teff}{\ensuremath{\mathit{T_{eff}}}}

\begin{document}

\title{A pulsar-atmosphere model for PSR 0656+14}
\author{Don A. Lloyd, Rosalba Perna, Patrick Slane, Fabrizio Nicastro, Lars Hernquist}
\affil{Harvard-Smithsonian Center for Astrophysics \\
60 Garden Street, Cambridge, MA 02138}
\authoraddr{dlloyd@cfa.harvard.edu}

\begin{abstract}

We present a pulsar-atmosphere (PA) model for modulated thermal X-ray
emission from cooling magnetized neutron stars.  The model synthesizes
the spectral properties of detailed stellar atmosphere calculations
with the non-uniform surface properties anticipated for isolated,
aging radio pulsars, and general relativistic effects on photon
trajectories.  We analyze the archival \textit{Chandra} observations
of the middle-aged radio pulsar PSR 0656+14 with the PA model and find
it is an excellent representation of the phase-averaged X-ray spectrum
for a broad range of polar effective temperature $T_{p}$ and column
density $N_{H}\simeq 10^{20}\ \mathrm{cm^{-2}}$.  The spectral fits
favor a sub-solar neutron star mass $M\lesssim1.0\ M_{\sun}$, a large
radius $R=15-16\ \mathrm{km}$ for a source distance $d\simeq190\
\mathrm{pc}$, while retaining consistency with theoretical neutron
star equations-of-state.  The modulated spectrum constrains the
angular displacement of the pulsar spin axis to $30\pm2\arcdeg$ with
respect to the line of sight.

\end{abstract}

\section{Introduction}

The discovery of modulated thermal X-rays in \textit{ROSAT}
observations of a handful of neutron stars (NSs) provided early
evidence that the surface properties of these sources are
inhomogeneous.  Most NSs are rotation powered, and emit predominantly
non-thermal X-rays generated exterior to the star in the pulsar
magnetosphere \citep{becker97,possenti02}.  Thermal X-rays from a few
exceptional sources originate at the stellar surface and can be
distinguished from magnetospheric contributions owing to their
intermediate age: these are sufficiently aged to have experienced a
decline in synchrotron activity, yet not so old that they are
altogether too faint to be seen.  The ``Three Musketeers,'' PSR
0656+14, PSR 1055-52 and PSR 0630+18 (Geminga) are relatively bright
and weakly absorbed middle-aged radio pulsars which are among the
earliest known thermal emitters (see \citealt{ogel95} for a summary of
\textit{ROSAT} data).  These sources remain excellent candidates for
evaluating NS cooling models and surface characteristics, and are
objects of continued interest in \textit{Chandra} and
\textit{XMM-Newton} observations.  \citet{becker97} tabulated the
short catalog of confirmed thermal NSs from the combined
\textit{ROSAT} and \textit{ASCA} data sets, which include the
relatively young Vela pulsar.  These X-ray spectra are found to be
poorly described by a single blackbody and require in addition either
a power-law or second blackbody component.  The secondary ``hard''
thermal component is invoked to bridge the energy range between the
``soft'' thermal and power-law contributions and broadens the spectral
distribution of the soft X-rays.  This spectral model may be
interpreted as thermal radiation from the entire stellar surface
(soft) seen in relief against small hot-spots or otherwise externally
heated polar regions (hard), plus the contribution from the
magnetosphere at sufficiently high energies.  In this picture, a
careful measurement of the soft thermal component may allow for direct
measurement of intrinsic stellar properties -- radius $R$,
characteristic temperature $T$, magnetic field and surface
composition.  This is quite challenging in practice because a
reasonable interpretation of the spectrum requires a self-consistent
model of the pulsar's synchrotron continuum, stellar viewing angles
and distance, in addition to the NS surface properties and the
associated thermal radiation.

The principal difficulty posed by the soft X-ray spectra of these
pulsars is in reconciling their modulated thermal emission and
estimated distance with a model for the surface radiation.  Owing in
part to modest instrumental sensitivity to the peak energies of their
relatively cool surface radiation $(0.1-1.0\ \mathrm{keV})$, measures
of temperature, the intervening absorbing column and overall
normalization (i.e., radius and distance) are mutually affected.
These uncertainties can be compounded by discrepancies as large as a
factor of $\sim4$ from different techniques for estimating the
distance to some sources.  The average spectrum alone is not
sufficient to deduce the temperature at the stellar surface, where
local anisotropies likely play a significant role in the fraction and
profile of the pulsed radiation; the energy dependent pulse profiles
and modulation index are functions of the apparent viewing geometry
and radiation beaming patterns, but are also dependent on the
assumptions of the model for the relevant pulsar mechanism.  In the
absence of perfect energy resolution of the detector, the strength of
the modulation that the observer infers also depends on the amount of
interstellar absorption \citep{page95,perna00}.

Models with two distinct thermal components provide acceptable fits to
the mean X-ray flux of these sources, while more complicated surface
temperature distributions also fare well.  \citet{page95} considered a
smoothly varying blackbody distribution on the stellar surface and
noted that a non-uniform temperature distribution implies increasing
pulsed fraction with energy.  The integration of such a dipolar
temperature blackbody distribution is well represented by that of a
single temperature within the limits of the PSPC data sets.  X-ray
pulsations are suppressed by gravitational lensing of the stellar
surface for a dipolar field regardless of the details of radiation
beaming.  More complicated field geometries can be invoked to explain
observed X-ray pulsation from the radio pulsars, but require
unacceptable tuning of surface parameters \citep{page96}.  The
blackbody temperature configurations generate a pulsed fraction which
increases with energy, and seems incapable of explaining either the
non-monotonic growth in pulsed flux seen for the middle-aged radio
pulsars observed with \textit{ROSAT} and \textit{Chandra}, or the
phase shift in pulse synchronization seen at $0.3-0.5\ \mathrm{keV}$
in \textit{ROSAT} observations.  These may be partially resolved by
beaming of radiation owing to intrinsic anisotropies to explain both
phase-resolved measures.

Radiation emergent from a stellar atmosphere is inherently anisotropic
regardless of the local temperature and magnetic field and, when
combined with a non-uniform temperature profile, is capable of
generating considerable pulsar modulation.  The earliest effort to
describe X-ray pulsations from NSs with magnetized plasma atmospheres
was due to \citet{pavlov94}.  These models characterized the pulsar
emission with small polar caps having uniform normal magnetic field
while the remainder of the stellar surface was assumed to be too cool
to contribute to the total flux.  \citet{zavlin95} refined this
approach and, in particular, generalized their calculation to include
an inhomogeneous surface temperature distribution and dipolar field
geometry.  A significant result of this effort was the discovery of a
``crossover'' energy at which the pulsed X-ray spectrum could
experience a phase-shift; this transition depends critically upon the
details of the ionization equilibrium of cool NS atmospheres, and on
the assumed surface temperature profiles and magnetic field geometry.
The dipole model was found to be generally consistent with the
\textit{ROSAT} observations of the middle-aged radio pulsars.
\citet{meyer94} made a detailed fit to the \textit{ROSAT} observations
of Geminga using partially ionized magnetic hydrogen atmosphere
models; these authors considered a simple model for Geminga in which
the temperature and magnetic field of the star were assumed
homogeneous, effectively a magnetic monopole.  The harder spectrum of
the plasma atmosphere was not alone able to fully describe the
\textit{ROSAT} PSPC data and required the secondary polar cap
component, albeit with much smaller effective emitting area than that
deduced from dual blackbody fits.

\citet{perna01} derived pulsed X-ray spectra $(0.1-1.0\ \mathrm{keV})$
of rotating NSs from the surface temperature profiles of \citet{hh98}
by integrating the spectrum of a toy hydrogen atmosphere model
\citep{hh_rcw103} at each point on the stellar surface.  As found in
the blackbody analysis, two thermal components were required to
reproduce the averaged \textit{ROSAT} spectra of the radio pulsars;
the phase resolved properties of these spectra (pulsed fraction and
pulse phase-lag with energy) were also reproduced by assuming the cap
had different beaming properties than the remainder of the surface.
More specifically, \citet{perna01} found that the behavior of the
pulsed fraction, as measured by \textit{ROSAT} PSPC, could be
reproduced with the combination of a ``pencil beam'' component
generated over the entire surface of the star, and a hotter ``fan''
beamed component, produced at the polar caps.

\subsection[PSR 0656+14 observations summarized]{PSR 0656+14}

X-ray emission from PSR 0656+14 was first detected with
\textit{EXOSAT} \citep{cordova89}; the suggested X-ray pulsations were
later confirmed when evidence of X-rays pulsed at the radio period
$P\sim 385\ \mathrm{ms}$ was found in the \textit{ROSAT} PSPC
observations described by \citet{finley92}; this spectrum was inferred
to be consistent with a blackbody $\Tbb\simeq9\times 10^{5}\
\mathrm{K}$ having normalization consistent with $R=10\ \mathrm{km}$
assuming a source distance $d=500\ \mathrm{pc}$.  An optical
counterpart was found in ESO observations $(m_{V}\sim25)$ as reported
by \citet{caraveo94}, and optical pulsations displaced from the radio
phase by 0.2 were recovered from later observations \citep{shearer97}.
At X-ray energies, the correlation between spectral hardness and
rotation phase suggests the pulsation originates in a non-uniform
surface temperature distribution.  \citet{anderson93} compare these
PSPC data to contemporary \textit{ROSAT} HRI observations $(0.15-2.0\
\mathrm{keV})$ and conclude from the modulated X-rays that the pulsed
emission is probably no harder than the unpulsed spectrum.
\citet{greiveldinger96} considered joint fits to the \textit{ROSAT}
and \textit{ASCA} data sets found that the spectrum was best fit by
two thermal components $T_{s}\simeq 8\times10^{5}\ \mathrm{K},
T_{h}\simeq 1.5\times10^{6}\ \mathrm{K}$ for $N_{H,20}=1.7$ (expressed
in units of $10^{20}\ \mathrm{cm^{-2}}$); the fit was significantly
improved by inclusion of a power-law to accommodate the hard X-ray
tail seen with \textit{ASCA}.  \citet{edelstein00} find that
simultaneous fits to \textit{EUVE} and \textit{ROSAT} observations
require both cooler temperatures for the soft thermal component
$\Tbb\simeq 3.6\times10^{5}\ \mathrm{K}$ and larger absorbing columns
$N_{H,20}=2.21$; continuing the fit to the thermal optical
measurements \citep{pavlov96,pavlov97} allow for larger
$\Tbb=(4.7-7.4)\times10^{5}\ \mathrm{K}$ and smaller
$N_{H,20}=(1.1-1.8)$, consistent with \citet{greiveldinger96}.  In
either case, the hard thermal component is $\Tbb\simeq (1.1-1.2)\times
10^{6}\ \mathrm{K}$.  \textit{Chandra} LETG observations
\citep{marshall02} find double blackbody fit parameters consistent
with those of \citet{greiveldinger96}.  Optical photometry performed
by \citet{kurt98} suggests the non-thermal optical flux has a spectral
index roughly consistent with that of the hard X-ray emission;
subsequent IR, optical and near UV measurements are consistent with a
harder spectrum and evolution of the spectral index
\citep{koptsevich01}.  \citet{pavlov02} also made an ACIS-S
observation in CC mode, and examined the joint fit with the earlier
grating data $(0.2-6.0)\ \mathrm{keV}$.  The resultant three component
(BB+BB+PL) model is likewise consistent with that of Greiveldinger et
al.; these authors find a fit to dual magnetic atmosphere models
questionable because of the large derived stellar radius, assuming the
source distance $d\simeq 500\ \mathrm{pc}$.  \textit{HST} observations
of PSR 0656+14 provide strong evidence for identification of the
optical counterpart \citep{mignani00}.  The proper motion of this
counterpart has been measured but, owing to the uncertainty in the
pulsar velocity, does not constrain the source distance to better than
a factor of $\sim 3$.  The lower bound likewise agrees with the
photometric estimate for this pulsar $(150-200\ \mathrm{pc})$ by
\citet{golden99}.

The energy dependence of the pulsed fraction of thermal X-rays has
been measured from \textit{Chandra} ACIS-S observations in the
continuous clocking mode \citep{pavlov02}.  The pulsed fraction does
not grow continuously with energy, but rather exhibits lower pulsation
from 0.5-1.0 keV when compared to either lower or higher energy
intervals; this property is a significant constraint on comprehensive
models for the X-ray emission.  The earlier {\it ROSAT} observations
show an increase in the pulsed fraction with energy, and is seemingly
inconsistent with any decrease in pulsation on this same interval.
Both datasets reveal an apparent phase shift in the pulsed thermal
X-rays although the magnitude of the phase displacement differs and
energies affected differ between the two observations ($\sim 0.8$ keV
in {\it ROSAT}, $\gtrsim 1.0$ keV in {\it Chandra}).
\citet{possenti96} evaluated the pulsed fraction predicted by the dual
temperature blackbody model and concluded from the PSPC data that the
measured $\sim 9\%$ on $(0.1-2.4)\ \mathrm{keV}$ could be reconciled
with the viewing geometry $\chi=30$ estimated by \citet{rankin93}, but
only for gravitational redshift $z\lesssim 0.15$.

In this article, we describe the synthesis of light element fully
ionized plasma atmosphere models of \citet{lloyd_methods} with
realistic NS temperature and field configurations derived from NS
crustal conductivities \citep{hh98}.  The resulting model yields
self-consistent pulse properties of the pulsar surface radiation and
provides substantial improvement over previous efforts to describe
pulsed thermal X-rays from NSs.  We compare the pulsar atmosphere
model to archival \textit{Chandra} data for PSR 0656+14, and examine
the extent to which our model provides an adequate solution to both
the average spectral envelope of this pulsar and the energy-dependent
pulsed fraction.  The pulsar model is capable of reproducing the
average X-ray spectrum without an explicit secondary thermal
component, but fails to reconcile the observed pulsed fraction.

We organize our discussion as follows: the pulsar-atmosphere (PA)
model is detailed in \S\ref{Sec:Model}.  The extraction and reduction
of the \textit{Chandra} LETG and ACIS-S observations is described in
\S\ref{Sec:Data}, followed by the results of our spectral analysis in
\S\ref{Sec:Fits}.

\section{Pulsar atmospheres}
\label{Sec:Model}
The surface distribution of effective temperature on a NS depends on
the local magnetic field strength and its inclination with respect to
the surface normal vector, and is prescribed by the conductive
properties of the magnetized neutron star crust
\citep[e.g.][]{greenstein83,hernquist84,hernquist84a,hh98}.  Both the
field and temperature distributions are parameterized by their values
at the magnetic pole.  For a body-centered dipolar magnetic field, the
surface field strength at magnetic colatitude $\theta_{p}$ is
\begin{equation}
\label{eq:Bdist}
B(\theta_{p}) = \frac{B_{p}}{2} \sqrt{1+3 \cos^{2}{\theta_{p}}}\ .
\end{equation}

The magnetic field vector at this point is inclined with respect to
the local normal (i.e. radial vector) by an angle $\psi$
\begin{equation}
\label{eq:cospsi}
\cos{\psi} = \frac{2 \cos{\theta_{p}}}{\sqrt{1+3
\cos^{2}{\theta_{p}}}}\ .
\end{equation}

Thermal conductivity in the magnetized NS crust varies with the
magnetic field density and direction, resulting in a smooth variation
of effective temperature from the magnetic pole to equator; for fields
$B_{p}\gtrsim10^{11}\ \mathrm{G}$, Heyl \& Hernquist (1998a) find:
\begin{equation}
\label{eq:Tdist}
\Teff(\theta_{p}) = T_{p}\left[\left(\frac{B}{B_{p}}\right)^{1/5}
\frac{4 \cos^{2}{\theta_{p}}} {1+3 \cos^{2}{\theta_{p}}}
\right]^{1/4}\ .
\end{equation}

We may optionally superimpose the radiation from heated polar caps on
the pulsar temperature distribution.  The cap dimension is described
by the angular radius $\theta_{c}$, and its intensity pattern is
modeled with a normal, uniform surface magnetic field $B_{p}$ and
constant effective temperature $T_{c}$ on the magnetic colatitude
interval $[0,\theta_{c}]$.  Radiation from the polar cap will
supersede that of the pulsar distribution for $\theta_{p}$ in this
range.  We simplify integration of the total surface emission in the
ordinary dipole field geometry by taking advantage of the axial
symmetry of the $B$ and $T$ distributions.  The radiation field at an
arbitrary point on the stellar surface can be mapped into the
equivalent point on a narrow strip of magnetic longitude extending
from the (magnetic) pole to equator which we discretize into $\sim20$
surface elements having uniquely defined local properties
$\vec{B},\Teff$ from equations (\ref{eq:Bdist}-\ref{eq:Tdist}).  Each
element is modeled using the computational methods described in
\citet{lloyd_methods} to derive the asymmetric thermal intensity
patterns for a hydrogen plasma atmosphere with these parameters.  The
model intensities $I_{E}(\vec{k})$ from all elements are tabulated for
interpolation of the radiation profile from an arbitrary point on the
NS surface.  The radiation field is fully symmetric about the magnetic
equator, including the two antipodal caps.

\subsection{Phase resolved model spectra}

The viewing geometry of the NS is defined by two angles: the angle
$\chi$ between the line of sight (LOS) and the pulsar spin axis
$\vec{P}$, and the relative angular displacement $\xi$ of the magnetic
dipole axis $\vec{B}_{dip}$ from $\vec{P}$.  Consequently, the angle
$\alpha$ between the magnetic dipole axis and the LOS varies with
phase angle $\gamma\equiv 2\pi t/P$:
\begin{equation}
\vec{B}_{dip} = (\sin{\chi}\cos{\xi}-\cos{\chi}\sin{\xi}\cos{\gamma},
-\sin{\xi}\sin{\gamma}, \cos{\alpha})
\end{equation}
where
\begin{equation}
\cos{\alpha} = \cos{\chi}\cos{\xi}+\sin{\chi}\sin{\xi}\cos{\gamma}.
\end{equation}
Any particular viewing geometry for which $\chi$ or $\xi=0$ is
degenerate, and fails to produce modulation of the surface emission in
the dipole model.  Results for choices of $\chi$ or $\xi$ which exceed
$\pi/2$ are equivalent to those for which the angle is subtracted from
$\pi$.

\begin{figure}
\includegraphics[width=6.0in]{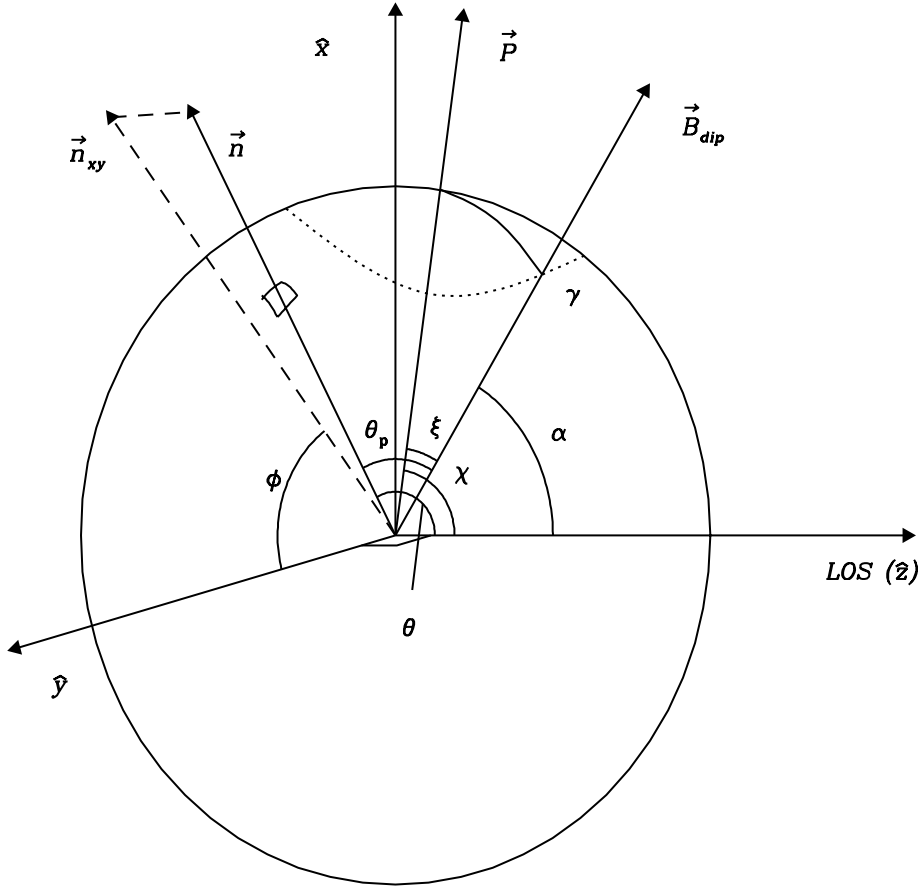}
% psrgeom.eps
\caption{ The global viewing geometry is specified by the pair of
angles $\chi,\xi$.  The dipole axis sweeps the dashed curve as the
star rotates though phase angle $\gamma$.  $\vec{P}$ is fixed and
coplanar with the line of sight $(-\hat{z})$ while $\vec{B_{dip}}$ is
coplanar with the LOS and spin axis $\vec{P}$ when $\gamma=0$ or
$\pi$.  Any surface element having normal $\hat{n}$ is referenced in
two coordinate systems: $(\alpha,\theta_{p})$ for some phase in the
pulsar coordinates, and by $(\theta,\phi)$ in the surface integration
coordinates.  $\phi$ is measured with respect to the projection of
$\hat{n}$ onto the $\hat{x}-\hat{y}$ plane (shown here as
$\hat{n}_{xy}$).}
\label{fig:psrgeom}
\end{figure}

Calculation of the phase-resolved flux proceeds from integration of
the intensity vectors coincident with a distant observer's LOS from a
large number of points on the stellar surface.  Two coordinate systems
are required, and computation of the model spectrum is facilitated by
the translation from \textit{viewing} or \textit{integration}
coordinates defined by a fixed LOS to NS \textit{surface} coordinates
defined with respect to the dipole axis.  This derivation proceeds in
Cartesian coordinates and assumes $\hat{l}=(0,0,-1)$ for the LOS.  To
deduce the correct intensity contribution from an arbitrary point on
the star, we require three \textit{surface} coordinates: the local
magnetic colatitude $\theta_{p}$, and the direction of the intensity
vector which is parallel to the LOS at distance and given by the pair
of angles $(\theta_{b},\phi_{b})$ in the NS frame.  The global
geometry is illustrated in Figure \ref{fig:psrgeom}.  The remainder of
our derivation describes recovery of this angular information from the
dipole pulsar model for arbitrary $\chi,\ \xi$ and $\gamma$.

The precise intensity contribution from each surface element varies
with rotational phase for a given viewing geometry, but the overall
configuration of surface elements remains fixed.  In a given phase,
each surface element is identified by a pair of polar
\textit{integration} coordinates $(\delta,\phi)$ where $\delta$ is the
angle between the intensity vector from the surface which intersects
the detector $[0,\pi/2]$ and the local normal vector, and the
circumpolar angle $\phi$ is on the interval $[0,2\pi]$.  The LOS is
the pole of this coordinate system.  The extent of the visible surface
is defined by $\delta=\pi/2$.  Note that, for relativistic stars,
photons with trajectories $\delta=\pi/2$ in these coordinates emerge
from surface latitudes $\theta>\pi/2$; in such cases the star is
self-lensing and has larger exposed radiative area than geometric area
for a hemisphere; for a non-relativistic star
$\theta\rightarrow\delta$.  To provide an absolute reference for
$\phi$, the prime meridian is defined such that the spin axis
$\vec{P}$ and LOS are coplanar with all normal vectors having $\phi=0$
regardless of phase angle; for $\gamma=0$, the magnetic dipole vector
is likewise coplanar with the spin and LOS.

\begin{figure}
\includegraphics[width=6.0in]{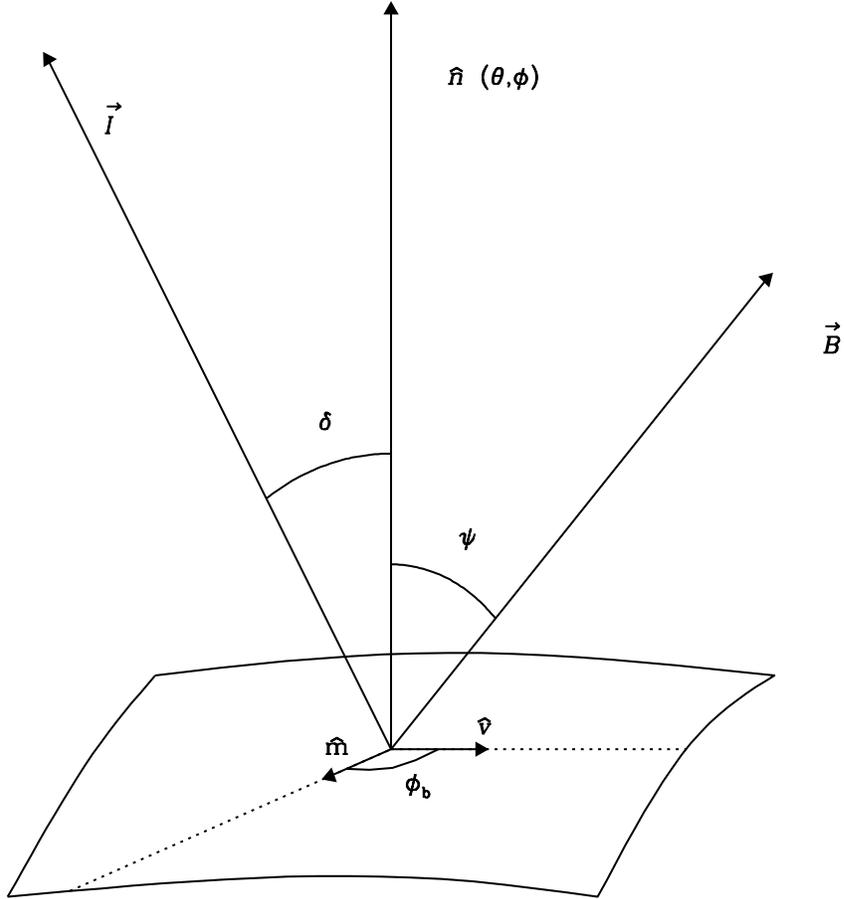}
% element.eps
\caption{ The surface element defined at the point referenced by
$\theta,\phi$ in the integration coordinates.  The intensity $\vec{I}$
intersects the detector at infinity and the angles $\delta$ and
$\theta$ are mutually defined by eqn. (\ref{eq:grray}).  Individual
surface elements are computed using the local $\vec{B}$ and $\Teff$
defined by the global (dipolar) magnetic field geometry.  The angular
intensity patterns are generally anisotropic, and the local azimuthal
angle $\phi_{b}$ is required for interpolation over the emission from
the table of computed elements.  }
\label{fig:element}
\end{figure}

Figure \ref{fig:element} summarizes the geometry at a given surface
element.  In our integration coordinates, the local normal vector at
the point $(\theta,\phi)$ on the stellar surface is
\begin{equation}
\hat{n} = (\sin{\theta}\cos{\phi}, \sin{\theta}\sin{\phi},
\cos{\theta})
\end{equation}
and has magnetic colatitude given by
\begin{equation}
\label{eq:costhetap}
\cos{\theta_{p}} = \hat{B}_{dip} \cdot \hat{n}
\end{equation}
The magnetic field at $\hat{n}$ is $\vec{B}=3
\cos{\theta_{p}}\hat{n}-\vec{B}_{dip}$ (cf. eqn \ref{eq:cospsi}).  In
the atmosphere description of the individual surface elements, the
local azimuthal reference is $\phi=0$ for coplanar $\hat{n},\hat{B}$.
The angle between the LOS and $\hat{n}-\vec{B}$ plane is
\begin{equation}
\cos{\phi_{b}} = \hat{m}\cdot\hat{v}
\end{equation}
where $\vec{v}=\hat{B}-\hat{n}\cos{\psi}$ and
$\hat{m}=-(\cos{\theta}\cos{\phi},\cos{\theta}\sin{\phi},-\sin{\theta})$
are the projections of $\vec{B}$ and the LOS onto the local tangent
plane.  $\phi_{b}$ is the azimuthal angle the intensity vector forms
with respect to the magnetic field axis and local normal vectors
\citep{lloyd_methods}.  Note that the intensity vector and local
normal are coplanar with the LOS.  To complete our definition of the
intensity vector, we measure the angle $\theta_{b}$ between the
desired intensity vector at the NS surface and the local normal, given
by the general relativistic ``ray-tracing'' integral \citep{page95}:
\begin{equation}
\label{eq:grray}
\theta_{b}(x) = \int_{0}^{1/2} x \left[\frac{1}{4}\left(
1-\frac{1}{R_{*}}\right) - \left(
1-\frac{2v}{R_{*}}\right)v^{2}x^{2}\right]^{-1/2} \d v
\end{equation}
where $x\equiv \sin\delta$, $R_{*}\equiv R/R_{S}$ and the
Schwarzschild radius $R_{S}\equiv 2GM/c^{2}$.

The intensity $I_{E}(\theta_{b},\phi_{b})$ from any integration
element is step-wise interpolated from the table of surface elements.
The analytic surface temperature distribution (\ref{eq:Tdist}) poorly
approximates the local flux for $\theta_{p}\gtrsim 85\arcdeg$; this
region contributes $\ll1\%$ of the total surface emission and
integration elements within this interval will be considered ``dark''
making no contribution to the total.  The phase resolved spectrum is
\begin{equation}
\label{eq:phasedflux}
F_{E}(\gamma) = \pi\left(\frac{R}{d}\right)^{2} \int_{0}^{2\pi}
\frac{\d\phi}{2\pi} \int_{0}^{1} \d(x^{2})\ I_{E}(\theta(x),\phi)
\end{equation}
The prefactors in eqn (\ref{eq:phasedflux}) normalize the total flux
to a distant source, where $R$ and $d$ are respectively the stellar
radius and source distance.  Relativistic self-lensing effects are
included in the intensity integration described above, while the
gravitational redshift will be accounted for when fitting the model to
the \textit{Chandra} data (\S\ref{Sec:Fits}).  The phase averaged
spectrum is
\begin{equation}
F_{E} = \frac{1}{2\pi} \int_{0}^{2\pi} \d\gamma F_{E}(\gamma)
\end{equation}
The average spectrum is typically calculated over $\sim20-30$ phases
of the pulsar rotation $\gamma=[0,\pi]$ while the integration of total
flux is performed by interpolating over on the order of 10,000-20,000
surface elements per phase.  The most useful representation of the
phase-resolved spectrum is the energy dependent pulsed fraction (PF)
\begin{equation}
\label{eq:PFEdfn}
PF_{E} = \frac{F_{E}^{max}-F_{E}^{min}}{F_{E}^{max}+F_{E}^{min}}
\end{equation}
In eqn (\ref{eq:PFEdfn}) the ``bright'' $(F_{E}^{max})$ and ``dim''
$(F_{E}^{min})$ phases are $\gamma=0$ and $\pi/2$ respectively which
are correct at X-ray energies for pulsar models in the absence of a
finite polar cap.  For cap models with finite temperature contrast,
the assignment of $\gamma$ for bright and dim may be reversed for
portions of the X-ray flux \citep{perna01}.  This is equivalent to a
phase change in the pulse synchronization for different spectral
components and is distinguished in these models by a negative $PF$.
For a realistic detector the pulsed fraction must be evaluated over a
finite band:
\begin{equation}
\label{eq:intPF}
PF_{E}(\Delta E) = \int_{E}^{E+\Delta E} PF_{E} \d E
\end{equation}
Instrumental efficiency and absorption in the ISM are energy dependent
and must therefore be included in the calculation of the integrated
pulsed fraction \citep{page95,perna00}.  As with gravitational
redshift of the spectrum, these effects will be incorporated into
spectral model from within the XSPEC software package.  The pulsed
fraction predicted in the PA model and others increases with large
$R$, owing to the diminished effect of self-lensing.

For a given polar magnetic field strength, the spectral model is
parameterized by the polar temperature of the pulsar distribution
$T_{p}$, the angular size $\theta_{c}$ and temperature $T_{c}$ of the
heated cap region, and the viewing angles $\chi$ and $\xi$.  The
pulsar atmosphere spectra are written in the FITS table format for use
in the XSPEC software package \citep{arnaud96}.  Polar caps may take
angular radii up to $13\arcdeg$ for temperatures
$\log{T_{c}}=5.7-6.5$, while the underlying pulsar pole temperature
takes values $\log{T_{p}}=5.4-6.0$.  The ``bright'' and ``dim'' phase
spectra are also included in the table to facilitate integration of
the pulsed fraction.  The pulsar viewing angles $\chi,\xi$ may assume
values from $0-90\arcdeg$.

\section{Data reduction}
\subsection{Low energy transmission grating spectrum}
\label{Sec:Data}

PSR 0656+14 was observed with the {\em Chandra} HRC-LETG configuration
(the combination of the High Resolution Camera imaging detector and
the Low Energy Transmission Grating - \citealt{brinkman00}) on 1999
November 28, with a total exposure time of 38 ks (ObsID 130).  We
retrieved the primary data products of this observation (i.e. standard
filtered events files, aspect solution, and bad-time/pixels filters)
from the public {\em Chandra} Data
Archive\footnote{http://asc.harvard.edu/cda/}, and extracted a
dispersed spectrum of the source with negative ande positive orders
co-added, and the background spectrum, using the {\em Ciao} (Chandra
Interactive Analysis of Observations) software (standard shape and
size extraction regions were used for both source and
background). Ancillary Response Functions (ARFs) for negative and
positive orders were also built with {\em Ciao}, and then co-added,
while for the redistribution matrix the HRC-LETG combination does not
allow us to resolve orders, so our final co-added spectrum contains
counts from all possible dispersed orders. However, contamination from
orders higher than the first, is only relevant at wavelengths longer
than $\sim 80$ \AA\ ($\sim 0.15$ keV), which we do not use in our
analysis. For the purpose of spectral analysis, we then built (with
{\em Ciao}) negative and positive 1st order HRC-LETG Ancillary
Response Functions (ARFs), and then co-added them. For the
Redistribution Matrix Function (RMF), instead, we used the ``canned''
1st-order HRC-LETG RMF distributed with the {\em Chandra} CALibration
Data Base (CALDB).

The total number of background subtracted counts in the source
dispersed spectrum, between 0.15 and 1.5 keV (the energy range used in
our spectral analysis) is 9230, corresponding to a signal to noise per
resolution element of $\Delta \lambda = 0.05$ \AA, of SN = 2.5 (6
counts per resolution element). Such a low number of counts per
resolution element does not allow us to search for narrow absorption
or emission features in the spectrum or to apply the $\chi^2$
statistics to fit models for the continuum emission to the unbinned
spectrum. We then grouped our spectrum allowing a minimum of 50 counts
per new grouped channel, and performed spectral fitting by using the
fitting package XSPEC on background subtracted data.

\subsection{ACIS-S3 CCD spectrum}

PSR~B0656+14 was observed for 25~ks on 15 December 2001, using the
ACIS detector in continuous-clocking (CC) mode (ObsID 2800). In this
mode, one imaging dimension is sacrificed in order to provide fast
read-out of the CCD. The resulting data consist of one row of events
for which each pixel value represents the sum of events from the
associated CCD column, with a time resolution of $\sim 3$~ms.  The
target was centered on the S3 chip, which is a backside-illuminated
device.  The energy resolution is $\Delta E/E \sim 0.1$ at 1~keV, and
this varies gradually with row number.

Source and background spectra were extracted from the data using the
{\tt dmextract} routine in {\em Ciao}.  Source events were extracted
from a region extending $\sim 1.8$~arcsec on either side of the
projected position of PSR~B0656+14, while background events were taken
from a $\sim 48$~arcsec region adjacent to the pulsar.  Spectra were
grouped to contain a minimum of 25 counts in each bin.

Calibration products for spectra obtained in CC-mode do not currently
exist.  As a result, we used spectral response and effective area
files created for a Timed-Exposure (TE) mode observation of
PSR~B0656+14 carried out on the same day, with the same pointing and
detector position configuration, as for the CC-mode observation.
Spectra from the TE-mode data were not used in these analyses because
the count rate for the pulsar is sufficiently high ($2.07 \pm 0.01
{\rm\ ct\ s}^{-1}$ in the CC-mode data) that pileup effects
significantly distort the spectrum.

\subsection{Timing analysis}

For investigation of the pulsed X-ray emission from PSR~B0656+14 we
extracted event times from a 1 x 1.5 arcsec$^2$ box which included
pixels for which the count rate was a factor of 10 or more higher than
the average for pixels outside the region of the pulsar
location. Event times were corrected from readout times to arrival
times by correcting for time offset effects from dither of the
spacecraft and any motions of the Science Instrument Module, following
the science analysis thread on CC-mode time corrections with {\em
Ciao}. These times were then corrected to the solar system barycenter
with the {\tt axbary} task, using the nominal coordinates of the
pulsar (RA$_{2000}$: 06:59:48.122, Dec$_{2000}$: +14:14:21.53).

\begin{figure}
\includegraphics[width=6.0in]{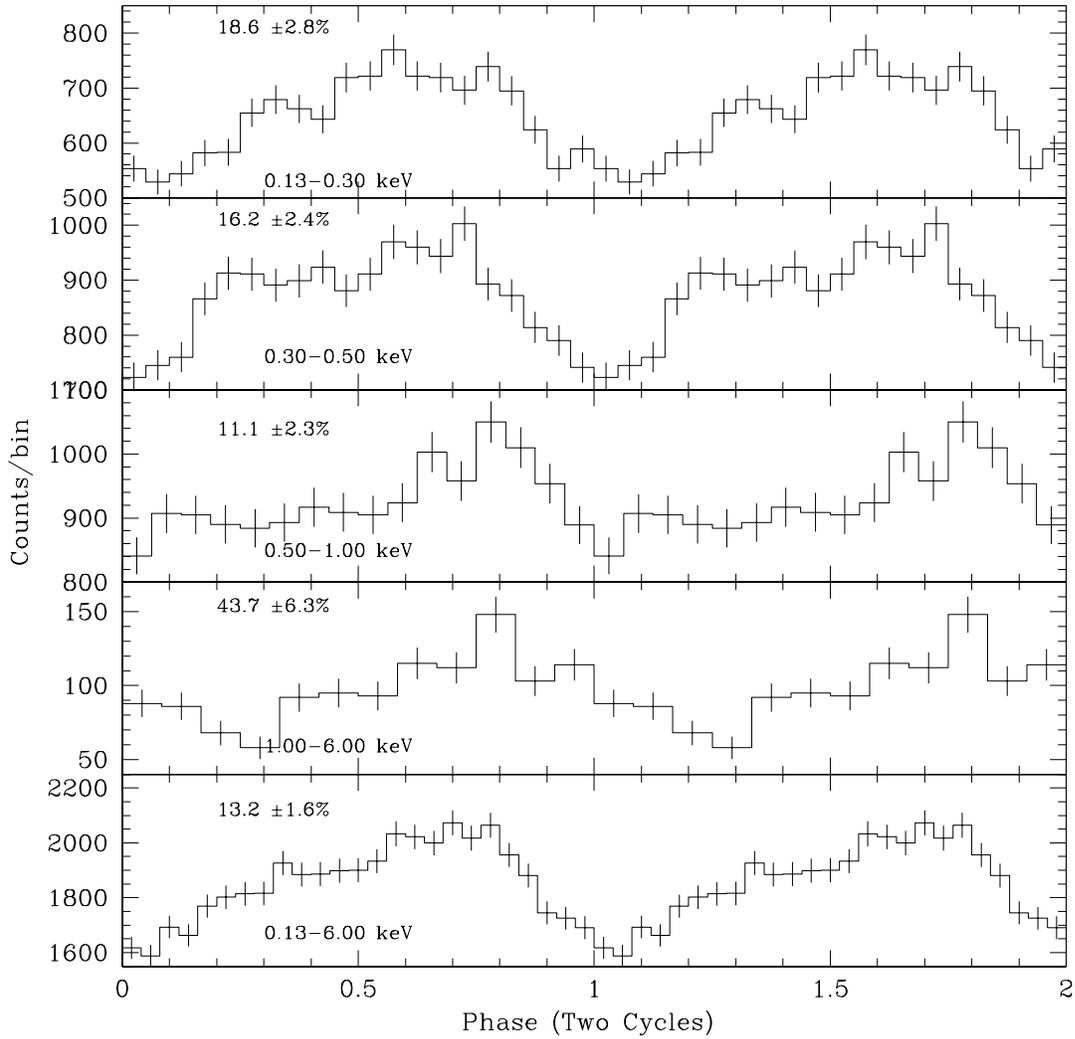}
% ltcrv.eps
\caption{ X-ray lightcurves for PSR 0656+14 accumulated with
\textit{Chandra} ACIS, integrated on representative energy
intervals. The pulsed fraction for each band is indicated in the upper
left corner of the panel.  The pulse profiles are irregular and are
not well described by the high degree of symmetry intrinsic to our
pulsar atmosphere model.  The energy intervals are the same as found
in \citet{pavlov02}.}
\label{fig:ltcrv}
\end{figure}

A search for periodicity around the expected frequency of
2.59805548~Hz, based on the ephemeris of \citet{chang99}, was carried
out using the $Z_n$ test \citep{buccheri83}.  The pulsations were
easily detected with $f = 2.598054 \pm 1.5 \times 10^{-6}$~Hz.
Absolute phases for each event were then calculated based on the above
ephemeris (see Table 1). Pulse profiles for selected energy ranges are
shown in Figure \ref{fig:ltcrv}, along with the measured pulsed
fraction for each profile, defined here as $p = (M-m)/(M+m)$ where $M$
is the maximum number of counts in the folded lightcurve and $m$ is
the minimum.

\begin{table}
\caption{PSR~B0656+14 Ephemeris}
\begin{center}
\begin{tabular}{lc}\\ \hline
Quantity & Value \\ \hline
Epoch (MJD) & 50701.000002341 \\
$f$ (Hz) & 2.5981054226747 \\
$\dot f ({\rm Hz\ s}^{-1})$ & $-3.71150 \times 10^{-13}$ \\
$\ddot f ({\rm Hz\ s}^{-2})$ & $8.33 \times 10^{-25}$ \\ \hline
\end{tabular}
\end{center}
\end{table}

\section{Spectral analysis}
\label{Sec:Fits}
Radio measurements of PSR 0656+14 suggest the pulsar has a magnetic
field strength $B\simeq 4.7\times 10^{12}\ \mathrm{G}$ in the dipole
braking model.  Its spectrum has been observed at optical and UV
energies, and \citet{ramana96} reported the possible detection of
$\gamma$-ray pulsations at the 385 ms rotation period in the EGRET
data.  Neither atomic (ionization) nor cyclotron features are evident
at any wavelength, and a direct measure of the surface magnetic field
is not possible.  For this analysis we assume that the surface polar
strength does not differ from that estimated from radio measurements,
consistent with the absence of either electron or proton cyclotron
absorption at X-ray energies.  Our analysis proceeds from the combined
LETG ($0.3-0.9\ \mathrm{keV}$) and ACIS-S ($0.4-6.0\ \mathrm{keV}$)
data sets.

A uniform (single component) blackbody spectrum will not produce
modulated flux, and both single temperature blackbody (BB) and
uniformly magnetized neutron star atmosphere spectra (ATM) also fail
to reproduce the phase-averaged \textit{Chandra} spectrum; these
elementary models can be discarded immediately.  The next simplest
alternative which does permit modulation is the composite of either
uniform thermal emission and a power-law (PL) component or a
double-blackbody (DBB) configuration with finite temperature contrast.
The ACIS spectrum is non-thermal above 2.0 keV; the DBB model must
also include the PL component to obtain a satisfactory fit.  We fixed
the power-law component to be consistent with the spectral fit to
optical measurements as described by \citet{kurt98}. The ATM+PL yields
a formally acceptable fit to these data, although the resultant
power-law amplitude is sufficiently small that temperature variation
is required to account for the $PF$ below 2.0 keV, even for 100\%
pulse index in the PL flux at these energies.  The extent to which the
non-thermal emission is pulsed is arbitrary in this model; however, no
evidence of pulsations above $2\ \mathrm{keV}$ was found in
\textit{RXTE} observations of this source \citep{chang99}, and we
assume the non-thermal flux is unpulsed at all energies in the
\textit{Chandra} data.  It is possible that finite pulsation of
non-thermal emission at energies $\lesssim 6.0\ \mathrm{keV}$ could
remain consistent with the \textit{RXTE} measurement.  The combined
BB+PL is a poor representation of these X-ray data, and we find that
although the DBB plus PL model provides an acceptable fit to the
average spectrum ($T_{1}=0.85\ \mathrm{MK}$, $T_{2}=1.5\ \mathrm{MK}$,
$A_{2}/A_{1}=3.4\times 10^{-2}$ \citealt{pavlov02}), this
configuration does not reproduce the observed pulsed fraction at
0.5-1.0 keV for any choice of viewing angles.

\begin{figure}
\includegraphics[width=6.0in]{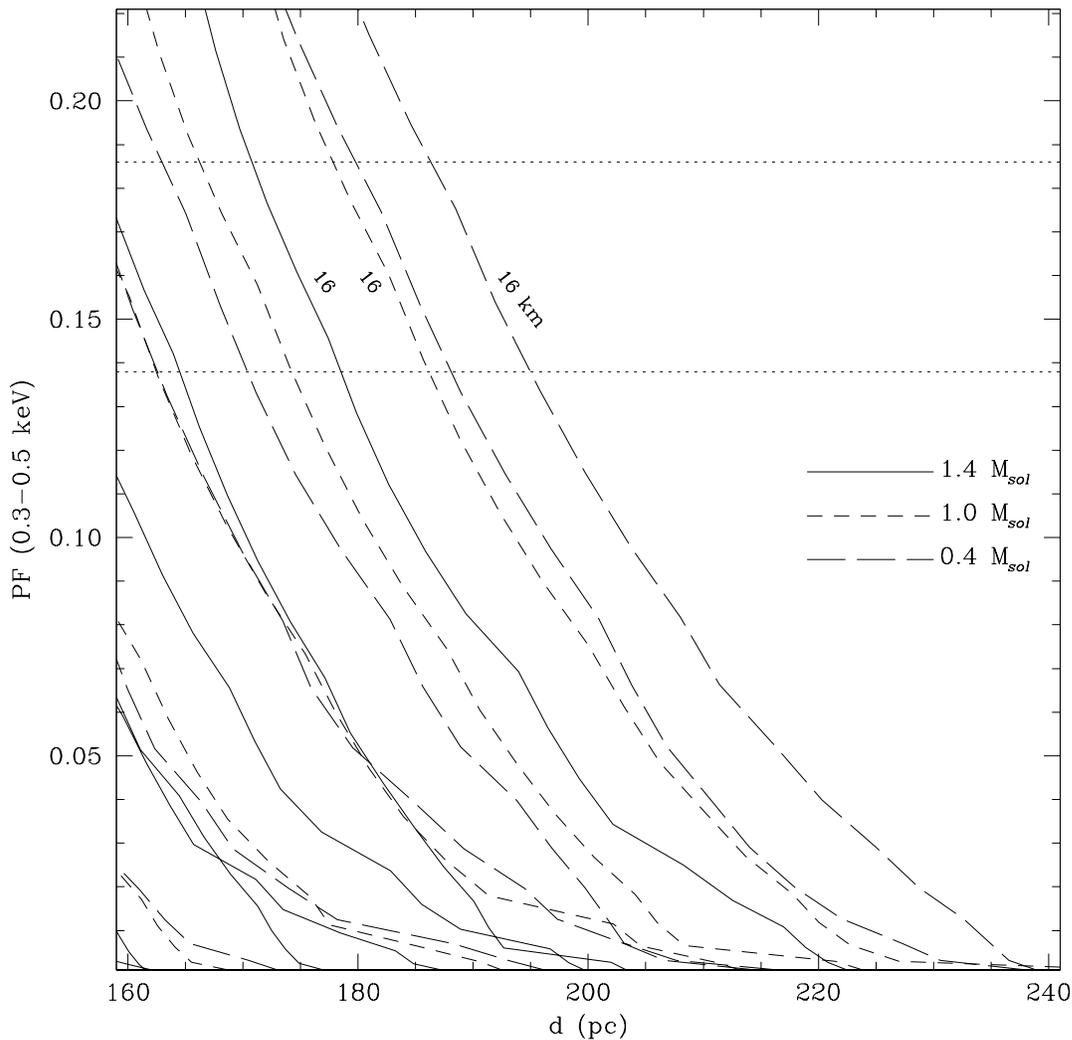}
% modlgrid.eps
\caption{ The grid of PA+PL models, organized by mass and radius.
Each curve is the locus of predicted $PF$ and $d$ for each choice of
$M,R$ and $\chi$.  16 km tracks for each mass are labeled, and curves
for lower radii (in 1 km increments) are displaced to lower $d$ for large inclinations
$\chi$.  For each curve, the viewing angle $\chi(=\xi)$ increases from
$\sim 2\arcdeg$ from the bottom of the figure to $\sim 35\arcdeg$ at
the top.  Well-defined spectral fits were obtained for $\gamma=1.5$
and $N_{H}=1.0-1.2\times 10^{20}\ \mathrm{cm^{-2}}$.  Distance and
$PF$ limits restrict further analysis to those models in the
rectangular area.  The best PA+PL models gives a 15-16 km stellar
radius for $d\simeq190\ \mathrm{pc}$, favors a sub-solar neutron star
mass and implies $\chi\simeq30\arcdeg$. }
\label{fig:modlgrid}
\end{figure}

We evaluated the combined \textit{Chandra} data sets with a grid of PA
model spectra for a grid of mass $M=(0.4,1.0,1.4) M_{\sun}$, radius
$R=10-16\ \mathrm{km}$ in steps of 1 km, and temperature
$\log{T_{p}}=5.4-6.0$ in steps of 0.1 dex, and we obtained good
spectral fits ($\chi^{2}/\mathit{dof}\lesssim 1.0$) with each pair of
$M,R$ for a range of viewing angle $\chi$ on the interval
$2-44\arcdeg$.  Here, we have assumed $\chi=\xi$ to reduce the
parameter space, which maximizes the pulsed fraction while demanding
the magnetic pole sweep through the line of sight.  For each pairing,
we assumed the non-thermal emission was well represented by a
power-law having index $\gamma=1.5$.  The spectral fits favor low
absorbing columns $N_{H,20}=1.0-1.2$.  From this grid, we selected
those models which generate the observed pulsed fraction from 0.3-0.5
keV, and maximized the source distance (Figure \ref{fig:modlgrid}).
The PA model provides a good fit to the observed average spectrum for
a range of $T_{p}=0.48-0.62\ \mathrm{MK}$ $(T_{p}^{\infty}=0.45-0.5\
\mathrm{MK})$ yielding a source distance $100\lesssim d \lesssim 250\
\mathrm{pc}$.  The thermal component $PF(0.3-0.5\ \mathrm{keV})
\lesssim 0.32$ over the grid, but also $ \lesssim 0.36$ above 1.0 keV.
The deficit of pulsed emission predicted by the PA model may be
accounted for by finite modulation of the non-thermal radiation at
these energies at the level of $\sim 70\%$.

\begin{figure}
\includegraphics[width=6.0in]{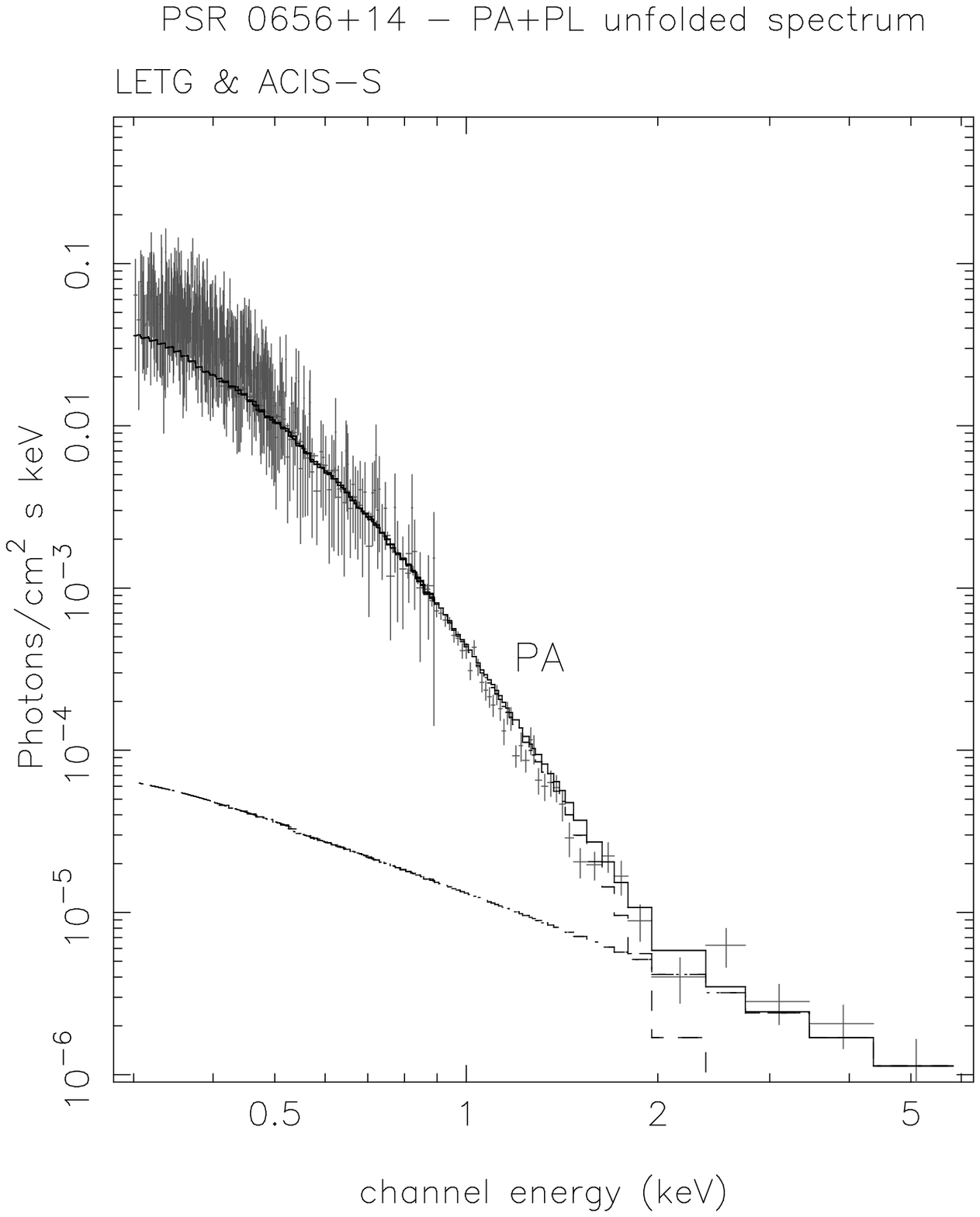}
% psr0656_pa_pl.eps
\caption{ The unfolded model and spectrum for $R=16$ km enclosing 0.4
$M_{\sun}$ at $d\simeq190\ \mathrm{pc}$.  These stellar parameters are
consistent with several equations-of-state \citep{lattimer01}.
Non-thermal emission dominates above 2 keV and is arbitrarily pulsed
in our model.}
\label{fig:paplmodl}
\end{figure}

\begin{figure}
\includegraphics[width=6.0in]{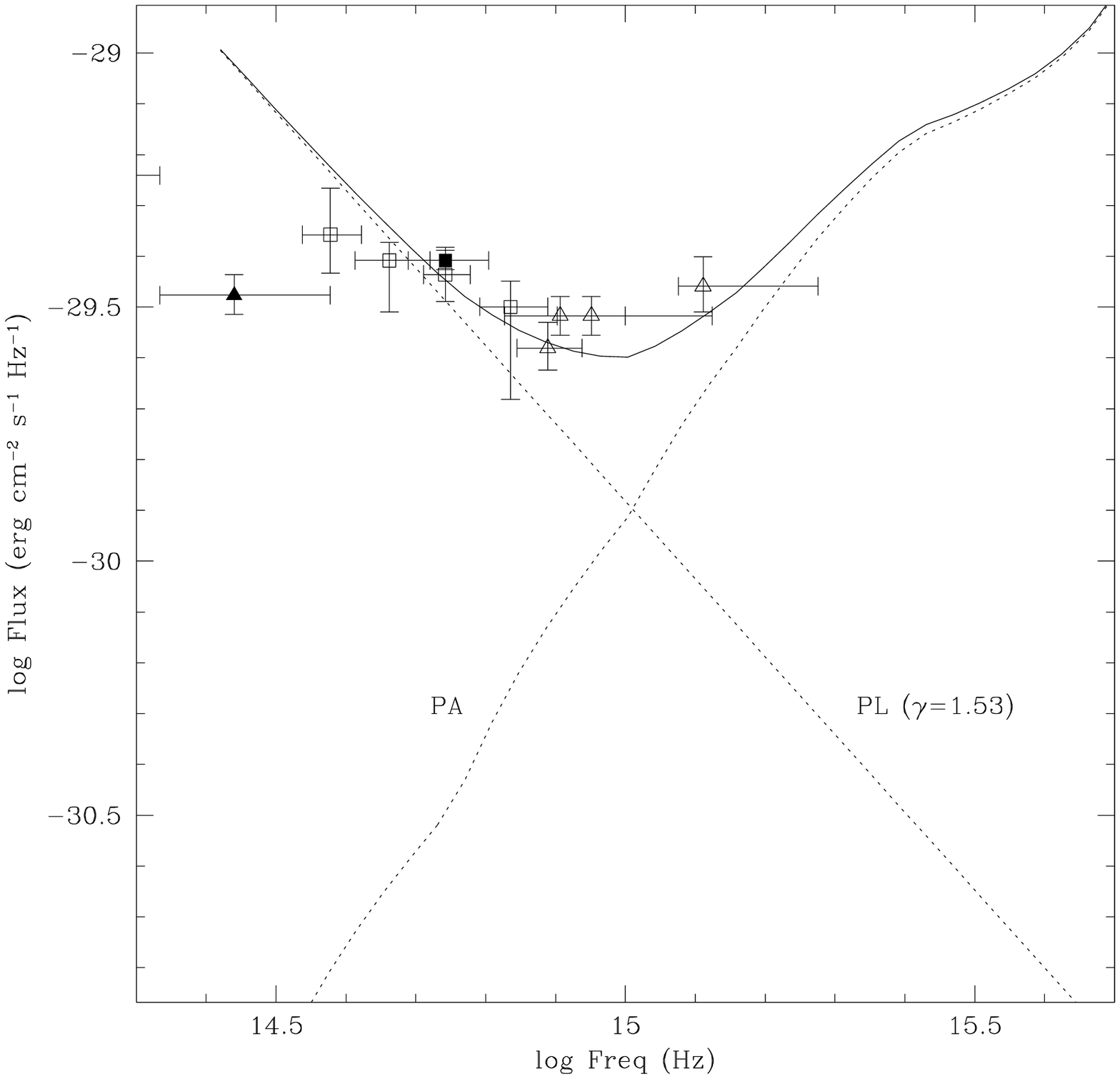}
% optical.eps
\caption{ The two component optical model for PSR 0656+14.  Data
points are those compiled in \citet{koptsevich01}.  The optical flux
is dominated by the non-thermal component, described here by the same
power-law used to model the hard X-rays in the {\it Chandra} data.
The thermal component has an approximately Rayleigh-Jeans spectrum and
does not over-produce optical flux.  The broad feature centered at
$\sim 7 \times 10^{15}\ \mathrm{Hz}$ is proton cyclotron absorption in
the hydrogen plasma. }
\label{fig:optical}
\end{figure}

The distance and pulse index constraints favor sub-solar NS masses and
large stellar radii.  To proceed, we consider in detail the PA model
results for a particular choice which satisfies our constraints,
$M=0.4 M_{\sun}$ and $R=16\ \mathrm{km}$.  We recovered $PF(0.3-0.5\
\mathrm{keV})$ within the measured uncertainties in our models for
$T_{p} \simeq 0.5\ \mathrm{MK}$ and $\chi=30\pm2\arcdeg$.  The
unfolded spectral model is illustrated in Figure \ref{fig:paplmodl}.
The PA model flux peaks at a lower energy compared to a uniform
surface at $T_{p}$ and has a broader spectral envelope owing to the
area-weighted average of the surface distribution, allowing for
excellent representation of the X-ray spectrum.  The PA component does
not over-estimate the optical flux of the pulsar, which has a
non-thermal spectrum (Figure \ref{fig:optical}).

For $\chi=\xi$, the ideal pulsed fraction predicted by the model
exhibits two trends with $R$ and $\chi$.  Spectral fits for fixed $R$
produce larger net pulsed fraction for increasing $\chi$ owing to
larger flux contrast as the bright magnetic pole is inclined from the
line of sight.  This has the incidental effect of requiring slightly
larger model normalization, favoring lower $d$ as the overall flux is
reduced on average.  Second, the magnitude of self-lensing is reduced
by either increasing the radius or reducing the stellar mass, thereby
increasing the pulsed fraction at all energies.  The effective
emitting area of the star grows with increasing $R^{\infty}=R(1+z)$,
and spectral fits to the pulsar spectrum accommodate greater source
distance.

\begin{figure}
\includegraphics[width=6.0in]{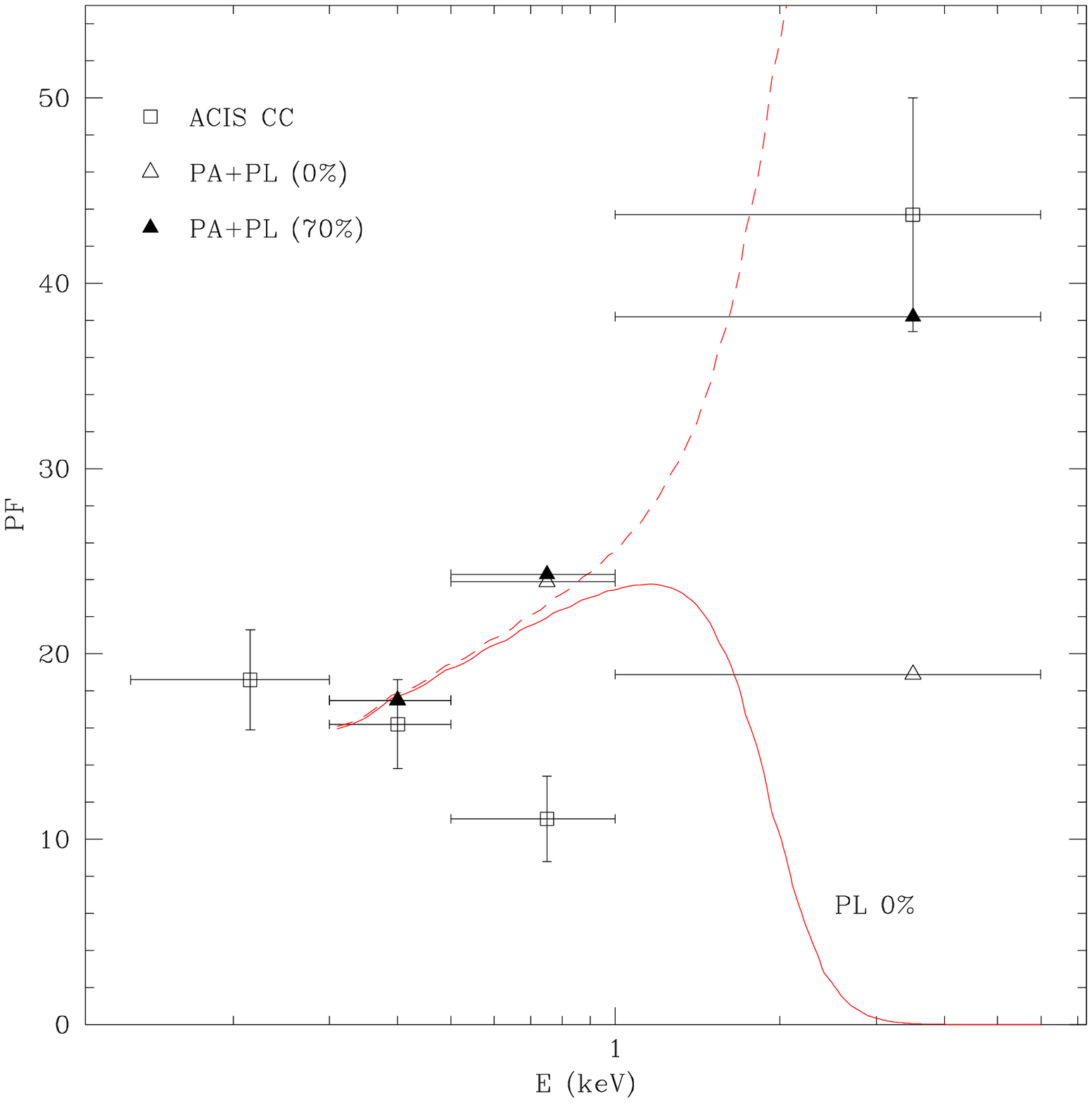}
% pf.eps
\caption{ The pulsed fraction for PSR 0656+14.  The measured net $PF$
integrated from 0.13-6.0 keV is 13.2\%.  The convolved model $PF$ for
the parameters of Figure \ref{fig:paplmodl} (open triangles) fails to
reproduce the pulsed fraction from 0.5-1.0 keV.  The solid curve
illustrates the thermal contribution to the pulsed fraction for the
same model.  Modulation of $\sim 70\%$ in the PL component is
sufficient to reproduce the measured $PF$ above 1 keV (filled
triangles), but does not lower the predicted pulsation from 0.5-1.0
keV; the dashed curve illustrates the $PF(E)$ for this model.
Calibration issues prevent using the spectrum below 0.3 keV, although
these data are expected to faithfully represent the pulsed fraction at
lower energies.  }
\label{fig:pf}
\end{figure}
The energy dependence of the pulsed fraction (Figure \ref{fig:pf}) is
an especially restrictive constraint on models for thermal X-ray
production for PSR 0656+14.  \citet{page95} demonstrated how
continuous variation of the surface temperature naturally produces a
$PF$ which increases monotonically with energy.  We attempted to model
the pulsed fraction at all energies by forcing finite polar cap
emission with temperature contrast $T_{c}>T_{p}$.  While formally
acceptable fits to the phase-averaged spectrum can be obtained from
these model combinations, the result in each case demands that the
``soft'' thermal component have both very low $T_{p}$ and large
stellar radius.  For example, we recovered $\theta_{c}\simeq2\arcdeg$
for $T_{p}=0.4\ \mathrm{MK}$ and $T_{c}=1.0\ \mathrm{MK}$ held
constant during the fit; by evaluating the $PF$ over a grid of
$(\chi$, $\xi)$ we found the most compatible values were
$\chi=80\arcdeg$, $\xi=50\arcdeg$.  The PA \& cap model is
nevertheless unacceptable as a comprehensive representation for the
pulsar thermal radiation: the radius expected for the relatively cool
$T_{p}$ component is 65 km for $d=200\ \mathrm{pc}$; alternatively,
the pulsar must have a distance of only 30 pc for the canonical radius
$R=10\ \mathrm{km}$.  Moreover, it is difficult to increase $T_{p}$
enough to substantially reduce $R/d$ without disturbing the relative
component amplitudes over 0.5-1.0 keV responsible for the ``dip'' in
pulsation on this interval; there is an effective limit $T_{c}\lesssim
1.3 \ \mathrm{MK}$, above which the heated cap becomes too small to
compensate for the observed growth of $PF$ with energy.

We also considered examples of discrete models having uniform
temperature and field over all but the antipodal caps, which had
finite temperature contrast.  As for the PA+cap models, the
temperatures and effective areas of the emitting regions were adjusted
to reproduce the average spectrum and provide a suitable crossover
energy.  Each of the discrete models requires either an unacceptably
large stellar radius ($R\gtrsim40\ \mathrm{km}$ for $d=500\
\mathrm{pc}$), or unreasonably short distance; and none has the
desired $PF$ evolution.  The magnetic field strength of the cap
$(B=0,~\mathrm{few}\ \times 10^{12}\ \mathrm{G})$ did not affect these
results.  We find the ``hard'' thermal component cannot have both
$\Teff$ greater than about $0.8\ \mathrm{MK}$ and sufficient amplitude
to induce the crossover on the required energy interval.

\section{Discussion}
\label{Sec:Discuss}

The amplitude of X-ray pulsations from PSR 0656+14 exclude uniform
temperature models for their origin, and require either a discrete or
continuous distribution of surface temperatures.  We have analyzed the
archival \textit{Chandra} observations of this middle-aged radio
pulsar with a variety of dual component thermal models and the PA
model of \S\ref{Sec:Model}.  We find that the phase-averaged spectrum
is well described by the combined PA+PL model for a broad range of
acceptable $M,R$ with the addition of a power-law component at
energies $\gtrsim 2\ \mathrm{keV}$.  By maximizing the source distance
and requiring that the resultant model reproduce the measured pulsed
fraction at $0.3-0.5\ \mathrm{keV}$ we find the PA interpretation
favors large NS radii $R\simeq15-16\ \mathrm{km}$, and stellar mass
$M\lesssim1.0~M_{\sun}$.  These results are compatible with several
equations-of-state for the neutron star core \citep{lattimer01} and
also indications from neutron star cooling analyses that the mass of
PSR 0656 is smaller than the canonical value $M=1.4 M_{\sun}$
\citep[e.g.][]{yak02,tsut02}.  For a particular choice of mass and
radius, $R=16\ \mathrm{km}, M=0.4\ M_{\sun}$ we recovered the viewing
angle $\chi=\xi\simeq 30\arcdeg$ and a source distance $d\simeq190\
\mathrm{pc}$.  Serious difficulties remain in reconciling the average
X-ray spectrum with the measured modulation.  In particular, the
apparent deficit of pulsed emission at $0.5-1.0\ \mathrm{keV}$ is not
readily explained with the PA model without the addition of a heated
polar cap; the discrete models we investigated fail to generate the
correct $PF$ evolution while models based on the pulsar type surface
can only reproduce the $PF$ for unreasonable $(R/d)$.  The model also
fails to generate the full measure of pulsation observed from
$1.0-6.0\ \mathrm{keV}$ without additional modulation of the
non-thermal flux; $\sim 70\%$ in the PL component would be sufficient
to close the gap above 1.0 keV.

The continuum spectrum from an iron envelope or from high metallicity
atmospheres bear some resemblance to a blackbody, although such metal
rich compositions have pronounced atomic line and ionization features
which should be evident at X-ray energies \citep{rr96}.  These
features are not found in the data; indeed, the (unmagnetized) Fe
models of \citet{gaensicke02} are a very poor representation of the
\textit{Chandra} observations, which favor the light element plasma
model we have described.  However, the presence of strong absorption
features in magnetized hydrogen plasma has previously been associated
with the formation of a soft X-ray crossover energy \citep{zavlin95}
above which the pulse profiles are phase shifted with respect to those
at longer wavelengths, and which also implies a reduced pulse contrast
(\ref{eq:intPF}).  For the magnetic field strength $4.7\times10^{12}\
\mathrm{G}$, the ground state binding energy of hydrogen is
approximately 0.25 keV \citep{potekh98}, below the usable lower limit
of these \textit{Chandra} data.  A magnetic field $B\simeq 10^{14}\
\mathrm{G}$ is required to increase the binding energy (and its
putative phase shift) to 0.5 keV; the X-ray data do not reveal this
feature and the pulsed fraction from 0.5-1.0 keV is not adequately
explained by photoionization.

The lightcurves predicted by the PA model are symmetric about the
$\gamma=0$ phase.  The observed X-ray lightcurves, however, show
irregular pulse profiles which are an indication of the fact that the
simple PA model is an oversimplification which lacks the necessary
variation in surface structure or beaming to generate the measured
$PF$ at all energies.

We found a restricted range of viewing angles $\chi\simeq30\arcdeg$
best represented the modulated thermal flux in our PA model.  This
result is reminiscent of estimations for $\chi$ from very different
wavelengths and emitting regions for this pulsar
\citep{malov90,rankin93}; \citet{rankin93} also deduced
$z\lesssim0.15$ consistent with the present results which favor
$z\simeq0.04$ or smaller.  \citet{harding98} have estimated the angles
$\chi$ and $\xi$ from a curvature radiation model to be smaller than
approximately $30\arcdeg$.  This is also consistent with the estimate
$\chi\sim 23.5-35\arcdeg$ derived by \citet{malov90} from three
measures of the radio pulse shapes.

Some models for $\gamma$-ray curvature radiation suggest a correlation
between the presence of $e^{\pm}$ currents and the extent to which
thermal surface X-rays can be modulated.  Models for the influence of
gamma-ray radiation mechanisms in the outer magnetosphere provide for
the production of X-ray power-law spectra and the formation of a
blanket of charge through which weakly modulated X-ray surface
radiation, and more strongly pulsed non-thermal polar cap radiation,
would be observed \citep{wang98}.  Pulsars without accelerators for
gamma-ray production would be dominated by the surface cooling
radiation at X-ray energies.  PSR 0656+14 is, at best, a weak source
of pulsed $\gamma$-rays \citep{ramana96}, which does not favor the
``hohlraum'' picture of \citet{wang98}; furthermore, the charge
blanket scenario provides an interpretation for strongly pulsed
non-thermal emission, but does not adequately explain the breadth of
the thermal X-rays.  The X-ray model of \citet{cheng99} describes the
pulsar spectrum as primarily thermal in nature and produced through
heating of the stellar surface by return currents of curvature
radiation pairs from the polar gap and outer gap.  Their model
provides a unified mechanism for describing both thermal and
non-thermal X-ray emissions from PSR 0656+14 for a viewing geometry
similar to our findings.

An additional problem arises from consideration of the extent to which
the atmospheric plasma can be heated externally while preserving the
thermal character of its net emission.  \citet{zane00} have calculated
spectra from model atmospheres heated externally (e.g. by charged
particle bombardment) --- the resultant X-ray spectra are essentially
unaffected by heating for modest particle fluence, although
amplification of the optical spectrum is apparent.  We have considered
the related problem of illuminated atmospheres, in which the surface
plasma is subject to an external radiation field of the form
$A\nu^{-\gamma}$.  The physics of illumination differs from that of
heating but models for each yield comparable results: heating of the
outermost strata of the atmosphere with concomitant enhancement of the
optical and near UV radiation.  The X-ray beaming functions (and hence
the $PF$) however are unaffected by the effects of illumination unless
the incident flux is so great that the atmosphere is completely
disrupted.

It is unlikely that the global magnetic field geometry for a given
pulsar is adequately described by the centered dipole which is assumed
in our pulsar-atmosphere model.  Contributions from higher multipole
moments, displacement of the dipole, stellar oblateness or localized
distortions of the field \citep{pageshib95} can impart additional
pulse components and erode the high degree of symmetry found in the PA
results.  Interaction between thermal radiation and charged clouds in
the magnetosphere \citep{wang98}, or with non-thermal currents may
also affect the X-ray lightcurves and deduced $PF$.

\acknowledgements
This work was supported in part by NASA contract NA58-39073 (PS).

\bibliographystyle{apj}
\bibliography{apj-jour,ns}

\end{document}